\documentclass[showpacs,aps,prl,groupedaddress]{revtex4-1}
\usepackage{amsmath,amssymb,graphicx}
\usepackage[usenames]{color}

\usepackage[top=1.5in, bottom=1.25in, left=1in, right=1in]{geometry}

\begin{document}

\title{Comment on the paper ``Quasi-particle approach for lattice Hamiltonians with large coordination numbers" 
by P. Navez, F. Queisser and R. Sch\"{u}tzhold - J. Phys. A: Math. Theor. 47 225004 (2014)}

\author{D.~Psiachos}
\email[ Email: ]{dpsiachos@gmail.com}

\begin{abstract}
This comment regards a central aspect of the referred-to paper, the issue of
convergence of the large coordination-number expansion. Perturbation expansions of
expressions containing a large number of parameters are generally invalid
due to the non-analyticity 
of the expanded expressions. I refer to 
recent work where these issues are analyzed and discussed in detail in relation to a benchmark 
example of a cluster model. As discussed therein, methods which are uncontrollable and for which their 
convergence is not foreseeable are not only useless but can mislead, particularly 
if models derived from them are used to interpret experiments. 
\end{abstract}

\maketitle

The paper~\cite{Navez} aims to study some properties of generalized lattice-based systems: from ones
with few parameters \textit{e.g.}  Heisenberg to more complex ones such as Bose-Hubbard (B-H) and Fermi-Hubbard (F-H).
The central methodology used is a perturbation expansion in large coordination number (Z$\gg$1) and it is applied to 
compute density matrices, correlation functions and then dispersion relations. The results of this expansion are compared with 
some exact solutions in 1D, outside the supposed range of validity but most importantly, not for sufficiently-complex models
with a large number of parameters such as the F-H and B-H models.

The models studied contain some or all of the following parameters: hopping, correlation, filling factor, Heisenberg exchange, and coordination number Z. 

However, it is well-known, and may easily be verified, that \textit{ad hoc} perturbation expansions do not in general work for 
expressions which are non-analytic as the
regions of convergence depend on the values taken by all the parameters present in the non-analytic portion. With 
an appropriate renormalization however, a new variable which does lead controllably to a convergent expansion may be defined, thereby
giving clear conditions for the convergence as regards the relationship of all the parameters amongst themselves. Just 
as an example, in my recent work on assessing series expansions for a cluster-model benchmark system~\cite{Psiachos},
in this case based on the two-band Anderson-Hubbard model,
I have shown that for expressions containing multiple parameters, which for anything but the most
trivial cases are most certainly non-analytic, a perturbation expansion must be performed in terms of a new, renormalized 
variable which combines the parameters in the non-analytic portion. Finding such a variable is very difficult, if not 
impossible, except for some simple situations. Only then can a region of convergence be defined. Otherwise, 
the region of convergence of the expansion will be dependent on the other parameters, 
in a way which is in practice unknown, thus rendering the expansion unreliable. This argument
equally holds for fixed order of expansion and increasing $Z$, \textit{e.g.} for an `asymptotic' expansion: it's not clear how the 
validity is impacted by the values of other parameters in relation to $Z$.  

In Eq. C2 of ~\cite{Navez} the authors present a comparison for a 1D (outside the supposed validity of the expansion) exact result - quantum Ising model - where the
result has been shown to agree only in some limiting forms for the parameters - not in general. For Eq. C1 (Heisenberg model),
it is \textit{possible} that owing to the few parameters involved in the model that an agreement with the
exact result is possible \textit{i.e.} if they are not present in the non-analytic portion. However in order to be useful for replacing
numerical calculations, such a method must be demonstrated to converge in a determinate fashion, in a 
foreseeable region of parameter space particularly for multi-parameter quantum-lattice models such as the B-H or F-H models treated
in Secs. 6-7. For that to be able to be achieved in all parameter space, the renormalized expansion 
variable must combine all those parameters found in the non-analytic portion of the full expression. The 
conclusions reached in Ref.~\cite{Psiachos} regarding convergence in expressions with
multiple parameters are sufficiently general so as to cover the work 
presented in the paper~\cite{Navez} even as they demonstrate their points using examples.

**EDIT** The same argument holds for paper~\onlinecite{NavezPRA}.


\end{document}